\documentclass[%
 reprint,
 amsmath,amssymb,
 aps,
]{revtex4-2}

\usepackage{graphicx}
\usepackage{bm}
\usepackage{amssymb}
\usepackage{amsmath}
\usepackage{epsfig}
\usepackage{braket}
\usepackage{epstopdf}
\usepackage{graphicx,epsfig}
\usepackage{mathrsfs}
\usepackage{dcolumn}
\usepackage{color}
\usepackage[dvipsnames]{xcolor}
\usepackage{natbib}
\usepackage{CJK}
\usepackage{bbm}
\usepackage{booktabs}
\usepackage{diagbox}

\begin{document}

\preprint{APS/123-QED}

\title{Robust translational invariance in Landau levels against lattice potentials and disorders}

\author{Bo Peng$^{1}$, Nilanjan Roy$^{1}$, Guangyue Ji$^{1,2}$, and Bo Yang$^{1,*}$}
\affiliation{$^1$Division of Physics and Applied Physics, Nanyang Technological University, Singapore 637371\\
$^2$Department of Physics, Temple University, Philadelphia, Pennsylvania, USA 19122\\
Corresponding author: yang.bo@ntu.edu.sg\\}
\begin{abstract}

We theoretically show that the continuous magnetic translational invariance within the Hilbert (sub-)space of a single Landau level (LL) can persist even in the presence of a superlattice electrostatic potential modulation, while such invariance is broken in the full real-space Hilbert space. This is due to the interplay of the superlattice constant and the fundamental length scale of the quantum Hall fluids. In particular for the lowest LL (LLL), when the spacing of superlattice is below the magnetic length, continuous magnetic translational symmetry is very robust. For the fractional quantum Hall phases, the continuous translational symmetry is preserved when the superlattice spacing is below the corresponding fundamental lengths which we can now quantitatively define, which is different from the length scale from the quantum metric.  Moreover, our analysis implies that the dynamics of the anyonic excitations can be robust against the short wavelength part of the  disorder potentials, and we discuss the related experimental ramifications.

\end{abstract}

\maketitle

\textit{Introduction--}
The integer and fractional quantum Hall effect(IQHE and FQHE)\cite{linxi1,linxi2} remain among the most studied examples of topological phases of matter\cite{Xiaogangwen1,topo2,topo3,phase1,phase2,phase3} with dissipationless transport and topologically protected Hall conductivity, which can be exposed as plateau in the transport measurement due to the disorder-induced localization of sparse quasi-particles\cite{linxi3}. Showing far-reaching implications in the fields of condensed matter physics\cite{implication2,implication3} and quantum information science\cite{implication4,implication5}, these QH states host quasi-particles with fractional charge and statistics\cite{statistics1, statistics2, statistics3}. In the presence of a strong magnetic field, the kinetic energy of the electron gas is quenched, leading to Landau levels, which are flat topological bands in two-dimensional systems. The rich physics of both integer and fractional quantum Hall effect fundamentally emerge from the rather special algebraic properties of the LLs\cite{algebraic1,algebraic2}. As the physically accessible Hilbert space at a large magnetic field and low temperature, the single LL truncates the whole quantum mechanical Hilbert space to a sub-space and provides a physical set-up where the non-commutative geometry naturally appears, thus the ``elementary particles" within it are no longer point particles\cite{non-commu1,non-commu2,non-commu3, elementaryparticles1, elementaryparticles2}.

As a vital concept in the study of Landau levels, magnetic translational symmetry modifies the conventional notion of translational invariance by incorporating the magnetic vector potential\cite{translation1,translation2}. This symmetry connects to topological phases and their phase stability and helps to understand anyonic dynamics in fractional quantum states\cite{topophases1,topophases2,topophases3stat,anyondynamics1}. In addition, it is related to the emergence of non-trivial band structures and edge states\cite{band1,band2,edge1,edge2}. Among these abundant physics of magnetic translational symmetry, the interplay between it and lattice translational symmetry is interesting, revealing crystalline order and phase transitions in strongly correlated systems\cite{latticeinterplay1,latticeinterplay2,topophases3stat,phasetrans,crystal1}. One interesting lattice description within a single Landau level is the von Neumann lattice (vNL) \cite{vNL1,vNL2,vNL3}. In a single LL, vNL states are a series of coherent states whose centers form a square lattice with a constant spacing, such that there is one flux quantum per unit cell. Notably, vNL states form a minimal complete basis even though they are \emph{not orthogonal} to each other\cite{vNL1,vNL2,vNL3,vNL-LL1}. Since each coherent state can be localized by a delta potential in real space (when LL mixing is ignored), the vNL of the coherent states can lead to a new perspective of superlattices within a topological band, with potential applications in quantum state engineering in LLs\cite{vNLengi,vNLengi2}.

While generally speaking, a lattice of delta potentials breaks the continuous translational symmetry down to discrete crystal symmetry in the full 2D Hilbert space, the completeness of the vNL of the coherent states suggests this could be a delicate matter for the Hilbert space of a single LL. Note that unlike the Hofstadter systems\cite{butterfly1,butterfly2,butterfly3} where the lattice potential (i.e. from the nucleus) is dominant over the LL energy spacing, here we are looking at the opposite limit where the lattice potential is \emph{smaller} than the LL spacing. Under a magnetic field, the uncertainty principle indicates that an electron can not be localized at one point but occupies a finite area on the order of the magnetic length. This implies that length scales smaller than the magnetic length cannot be physically resolved. Thus, it is interesting to see if the continuous translational symmetry can be preserved even in the presence of a lattice of one-body delta potentials with a small lattice constant. 

In this paper, we show the robust translation invariance in continuous QH systems against a superlattice of local potential and disorders when the length scale of such external electrostatic potential becomes comparable or smaller than the magnetic length. We investigate the behavior of local potential lattices confined to the Hilbert space of a single Landau level (LL) and show that the bandwidth decreases dramatically as lattice spacing decreases below the magnetic length. This implies almost exact continuous magnetic translational invariance in the presence of a discrete lattice potential. It allows us to define such a special type of superlattice Hamiltonian, the von Neumann Hamiltonian, that obeys the approximately continuous translational symmetry and determines the fundamental length of LLs, which we show to be different for different LLs. In addition, we generalize this concept to the interacting system of the fractional quantum Hall (FQH) states. Using the real space density oscillation profiles of the quasiholes\cite{guangyue}, we find that the fundamental length scale of the interacting conformal Hilbert spaces (spanned by the ground states and quasihole states of a particular FQH phase\cite{CHS1, CHS2, CHSspan, CHSspan2}) is significantly larger. This implies the properties of the anyonic charge excitations (such as the bandwidth and braiding) of the FQH states are \emph{more robust against disorder potentials} that are smaller than the incompressibility gap. The interplay between the superlattice spacing and the magnetic length gives insights into the dynamical properties of the elementary particles in the periodic electrostatic potential in the presence of a strong magnetic field and interaction. This also leads to a better understanding of the impurity effects in such systems in the experiments.

\textit{Lattice potential in the LLL--} We start our discussion with the coherent state in a single LL on the disk and introduce the useful notations. The kinetic energy of electrons in a magnetic field can be described by a Hamiltonian: $H=\frac{1}{2m}(\textbf{p}-e\textbf{A})^2$. With the magnetic length $\ell_B=\sqrt{\hbar/{eB}}$, two sets of spatial coordinates, cyclotron and guiding center coordinates are defined as: $\tilde{R}^a=-\epsilon^{ab}\ell_B^2(p_b-eA_b)$, $\hat{R}^a=r^a-\tilde{R}^a$, respectively, where $\epsilon^{ab}$ is the antisymmetric tensor and  Einstein’s summation convention is adopted. This allows us to define two sets of decoupled ladder operators $\hat{b}^\dagger=(\hat{R}^x+i \hat{R}^y)/\sqrt{2}\ell_B$ and $\hat{a}^\dagger=(\tilde{R}^x+i \tilde{R}^y)/\sqrt{2}\ell_B$, with $[\hat b,\hat b^\dagger]=[\hat a,\hat a^\dagger]=1,[\hat b,\hat a]=0$.  The commutation between the two sets of coordinates indicates that eigenstates of such Hamiltonian can be described by two indices, the LL index and the angular momentum index, as $\ket {N,n}=\frac{1}{\sqrt{N!n!}} (\hat{a}^{\dagger})^N (\hat{b}^{\dagger})^{n}\ket{0}$. The coherent state located at the origin in the LLL is defined by $\hat{a}\ket{0}=\hat{b}\ket{0}=0$, and its explicit wave function in the symmetric gauge is $\braket{z,z^*|0}=\frac{1}{\sqrt{2\pi}}e^{-\frac{1}{4}zz^*}$  with $z=x+iy$. All other coherent states in the LLL can be generated by acting the magnetic translation operator $T(X) \equiv e^{i\hat{R^a}X_a}$ on $\vert 0 \rangle$ as follows:

\begin{equation}
    \ket{X}=T(X)\ket{0}= e^{i\hat{R^a}X_a}\ket{0},
\end{equation}
whose centre is at $(X_y,-X_x)$ in real space.

In the presence of a lattice electrostatic potential, the energy degeneracy within a single LL will be split, and the LL will get a finite bandwidth. In this work, we model the lattice potential as:
\begin{equation}
        H =qU\sum_{x_0,y_0}\delta (x-x_0)\delta (y-y_0),
\end{equation}
in which $(x_0,y_0)$ are the positions of the delta potentials, forming a lattice in real space, and $q$ is the charge of the particle. In experiments, applying a voltage $U$ to the tip of the needle can be used to implement the delta potentials\cite{tip1,tip2}, thus $qU$ is the energy unit of local potential lattice, whose scale is much smaller than the gap between Landau levels, and we set $qU$ as unity hereafter. 

A Hamiltonian given by Eq.(2) clearly breaks continuous translational symmetry in real space. However, within the Hilbert space of a single LL, the effect of such a Hamiltonian is more subtle. The breaking of the translational symmetry can be quantified by the variational bandwidth from the change of the variational energy of $|X\rangle$, when the position $X$ is varied. For the square lattice whose delta potentials are at $(at,bt)$, where $a$ and $b$ are taken as integers and $t$ is the spacing of such lattice, this Hamiltonian gives the energy bandwidth for the single-particle states in the LLL(as shown in Fig.~\ref{fig1}(b)) that can be computed analytically\cite{supple}:
\begin{equation}
    \Delta E= \frac{2}{\pi t^2} \theta\left[2,0,e^{-\frac{8\pi^2}{t^2}}\right]\cdot \theta\left[3,0,e^{-\frac{8\pi^2}{t^2}}\right],
\end{equation}
where the elliptic theta functions are given by $\theta[2,u,q]=2q^{\frac{1}{4}}\sum_{n=0}^{\infty}q^{n(n+1)}cos((2n+1)u)$, and $ \theta[3,u,q]=1+\sum_{n=1}^{\infty} q^{n^{2}}\cos(2nu)$.

At the length scales significantly smaller than the magnetic length, the bandwidth of the local potential lattice(i.e. Eq.(3)) dramatically shrinks as the spacing decreases, even faster than exponential. This implies at small spacing, the Hamiltonian is very close to the identity matrix, thus the variational bandwidth will converge to the exact bandwidth from exact diagonalization of the Hamiltonian using the single particle states. This has been numerically verified both on the disk and torus geometry\cite{supple}. Unless otherwise stated, the variational bandwidth will be used for the rest of the work for the superlattice of delta potentials. It is more advantageous compared to exact diagonalization because of the analytic expression that we can obtain in the LLL. More crucially, exact diagonalization has strong finite size effect both on disk and torus geometry for the quasiholes from the fractional quantum Hall states, and for small lattice spacing the energy bandwidth can be more accurately computed by varying the position of the quasiholes.

\textit{The vNL Hamiltonian--} In analogy to the same lattice structure and completeness of vNL in the LLL, we now also define the vNL Hamiltonian within LLL as follows:
\begin{equation}
H_{_{vNL}}\!\!=\sum_{X}\ket{X}\bra{X}=2\pi P_{_{LLL}}\!\sum_{x_0,y_0}\delta(x-x_0)\delta(y-y_0)P_{_{LLL}},
\end{equation}
for lattice constants that are smaller than the magnetic length $\ell_B$, $P_{_{LLL}}$ is the projection operator to the LLL. With this definition, vNL Hamiltonian gives a nearly flat band with a narrow bandwidth smaller than $2.1\times 10^{-8}$.

Since the vNL Hamiltonian above is defined for a particular Hilbert space of the LLL, we can also generalize to higher LLs for integer quantum Hall effect. Moreover we can also generalize to other Hilbert spaces corresponding to the fractional quantum Hall phases. These are the so-called conformal Hilbert space (CHS)\cite{conformal1, conformal2,CHS1,CHS2}, among which a single LL is a special case. Each CHS is spanned by the ground state and the quasihole states of a particular topological phase. Using the CHS of the Laughlin $\nu=1/(2n+1)$ phase as an example, the model Hamiltonian is a sum of short-range two-body interactions \cite{pseudopotential} and at $\nu=1/(2n+1)$, the CHS is the null space of the model Hamiltonian spanned by the Laughlin ground state, and quasiholes, which are ``elementary particles'' within the CHS analogous to IQHE holes within a single LL, forming a dispersionless ``many-body flat band" with the model Hamiltonian. Within CHS, there can be continuous translational symmetry up to very good approximation (as evidenced by the very small single particle bandwidth) even in the presence of the delta potentials that explicitly breaks such symmetry in full Hilbert space. Similar to the LLL, the vNL Hamiltonian within CHS can be defined as a local potential lattice with discrete translational symmetry in real space, which gives a nearly flat band, with the quasihole bandwidth remaining below the same cut-off of $2.1×10^{-8}$ from the LLL.

\begin{figure}[!h]
\centering
\includegraphics[width=1.0 \linewidth]{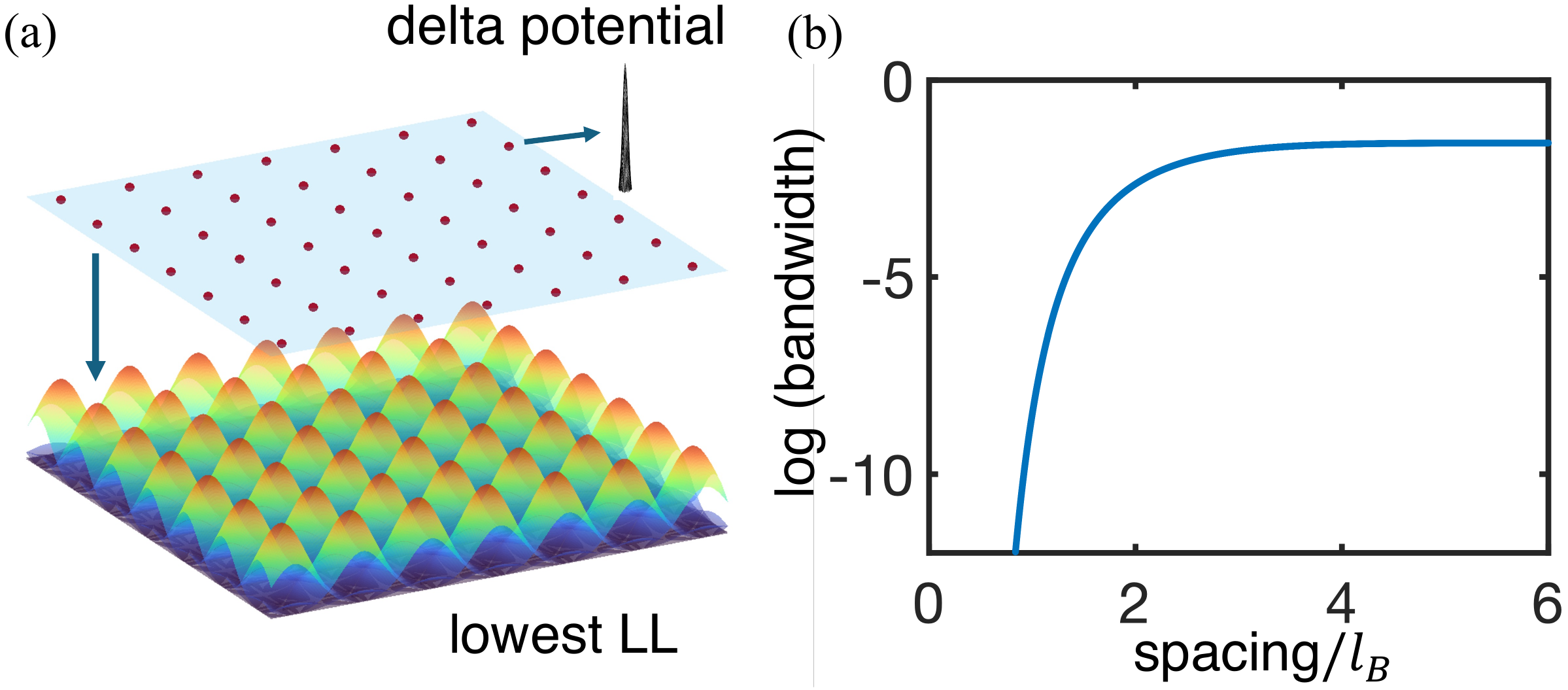}
\caption{(a). Schematic representation of a lattice of local delta potential square lattice in real space; the projection of such potentials into the LLL gives the equivalent lattice of projection operators, each represents a single particle coherent state. (b). Logarithm of bandwidth of square delta potential lattice on the disk in the LLL, as a function of the delta potential lattice constant.}
\label{fig1}
\end{figure}

\textit{Fundamental length scales of conformal Hilbert spaces--}
The vNL Hamiltonians we defined for a single LL give us a fundamental length scale $a_{LL}$, so that the magnetic translational invariance is hardly broken when the lattice constant $a\le a_{LL}$. Naturally, the fundamental length for LLL is $a_{LLL}=\ell_B$. It is important to note that such robustness comes from the truncation of Hilbert space. While different LLs can be physically the same in many ways, their fundamental lengths are different. With the bandwidth $2.1\times10^{-8}$ as the benchmark, the fundamental length scales for higher LLs on the disk are shown in Fig.~\ref{fig2}(a), in which the higher LL is, the smaller the fundamental length is. The fundamental length as a function of the Landau level index is best captured by the empirical expression $\tanh(a/N^b)$, with fitted parameters $a=1.277$ and $b=0.34$, shown in Fig.~\ref{fig2}(e). The empirical equation is motivated by the fundamental length vanishing in infinitely high LL, since with the full Hilbert space (all LLs included), any superlattice potential will break translational symmetry and lead to a large bandwidth no matter how small the lattice constant is.

It is interesting to note that different LLs are also distinguished by another length scale: the trace of the gauge invariant quantum metric given by $l_{qm}= \sqrt{2N+1}\l_B$ \cite{metric1,metric2,metric3,metric4,metric5}, which increases with the LL index. The contrasting tendency of the two length scales reveals that as LL increases, size of the maximal localized Wannier orbital gets larger but the length resolution of the corresponding CHS becomes higher. The fundamental length defined in this work is more relevant to physically measurable quantities related to the density profile of a single particle state in a single LL, and we will discuss in more details later on. 

For many-body CHS that are subspaces of the LLL, we can use their corresponding vNL Hamiltonians to characterize the fundamental length scales of the corresponding many-body topological phases. 

The computation of the bandwidth of the anyons within each CHS with respect to the lattice Hamiltonian requires knowledge of the real space density distribution of the quasiholes. This is highly non-trivial for interacting systems. We use the results from Ref.~\cite{guangyue}, which so far gives the most accurate real-space density distribution $\rho(r)$ in the thermodynamic limit. For the Laughlin states at filling factor $\nu=1/(2n+1)$, we now look at the bandwidth of a single Laughlin-1/3 quasihole in the presence of a superlattice of delta potentials. Due to the limitation of numerical accuracy, our resolution of the bandwidth is only up to the order of $10^{-5}$, thus we need to fit the relationship between the bandwidth and superlattice spacing for the extrapolation to a smaller bandwidth of $2.1\times10^{-8}$. In this case, for Laughlin-1/3 state, the lattice spacing of vNL Hamiltonian is $a\le a_{1/3}\sim 1.7\ell_B$ as shown in Fig.~\ref{fig2}(b) given by the blue crosses. We conjecture the fundamental length of $a_{1/3}$ should be $\sqrt{3}\ell_B$with such bandwidth benchmark. This obviously agrees with the composite fermion picture where the effective magnetic field for the composite fermions (one electron with two attached fluxes) at $\nu=1/3$ is one-third of the real magnetic field for the electrons, implying that the size ratio of Laughlin-1/3 quasiholes and IQH holes is $\sqrt{3}$. 

\begin{figure}[!h]
\centering
\includegraphics[width=1.0 \linewidth]{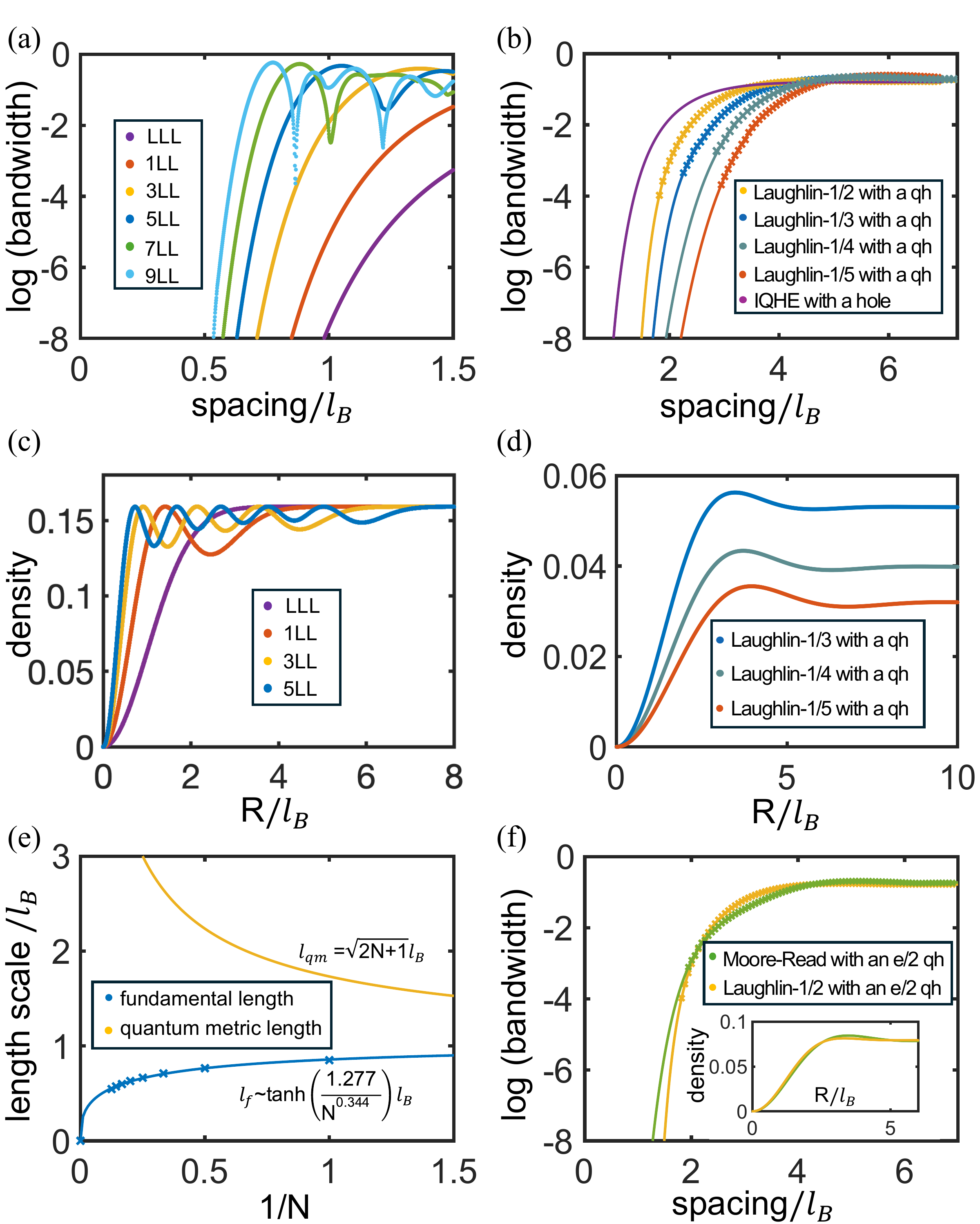}
\caption{The bandwidth in logarithm scale of a single (quasi-)hole as a function of the superlattice constant of (a) in different LLs; (b). for different Laughlin states; crosses here are numerical data computed from density profile, and lines are results from fitting. The violet line is the bandwidth of IQHE hole for comparison. (c). The damped oscillation of the electron density of a hole in different LLs. (d). The damped oscillation of the electron density of a quasihole in different Laughlin phases. (e). The fundamental lengths (blue crosses and line, fitting with $tanh$ function, with the $R^2$ coefficient as 0.9997.) and the exact trace of the quantum metric, as a function of the LL index. (f) Same as (b). Yellow crosses and line are about Laughlin-1/2 state with an $e/2$ quasi-hole, and green crosses and line are about Moore-Read state with an $e/2$ quasi-hole. The inset figure is about the density profiles of the corresponding quasiholes.}
\label{fig2}
\end{figure}

The logarithms of bandwidths as a function of lattice spacing for Laughlin states are shown in Fig.~\ref{fig2}(b)  with different filling factors after the modification. The yellow, blue, dark cyan, and red crosses are about the Laughlin-1/2, Laughlin-1/3, Laughlin-1/4, and Laughlin-1/5, with a quasihole, respectively. For comparison, the violet points are from the hole bandwidth of the LLL IQH state. The fundamental length scale obtained numerically are around $1.5\ell_B$, $1.7\ell_B$, $2.0\ell_B$ as well as $2.2\ell_B$ for  Laughlin1/2, Laughlin1/3, Laughlin-1/4 and Laughlin1/5 state respectively, very close to the conjectured values $\sqrt{2}\ell_B$, $\sqrt{3}\ell_B$, $\sqrt{4}\ell_B$ and $\sqrt{5}\ell_B$, consistent with the composite fermion theory and the hierarchical nature of the CHS. For the Abelian Laughlin states, this fundamental length increases with decreasing filling factors, intuitively characterizing the size of the corresponding quasiholes in real space. This is also consistent with the decreasing of the fundamental length in higher LLs: both the electrons/holes of IQH in different Landau levels and the Laughlin quasiholes have a Gaussian decay part and the oscillation part in the density profile (i.e. well fitted by the damped oscillation model in Ref.\cite{guangyue}), as shown in Fig.~\ref{fig2}(c) and~\ref{fig2}(d). In higher LL or larger filling factor for Laughlin quasiholes, both the decay length and the oscillation length decreases, illustrating the physical significance of the fundamental length scale for the quantum Hall fluids. 

It is also interesting to look at the non-Abelian Moore-Read states; they span the null space of a short-range three-body interaction. Insertion of a single flux creates two quasiholes, each carrying a charge of $e/4$. There is no single-quasihole subspace, and we only have the real space density profile of the $e/2$ quasihole, from two $e/4$ Moore-Read quasiholes stacked on top of each other in real space. This $e/2$ quasihole forms a well-defined Abelian subspace, and the green crosses in Fig.~\ref{fig2}(f) show the bandwidth of superlattice within such a subspace. We also give the extrapolation with fitting, extracting the critical spacing for bandwidth $2.1\times10^{-8}$ as $1.29\ell_B$. At the filling factor $1/2$, this value for the non-abelian Moore-Read state is significantly different from $\sqrt{2}\ell_B$, the fundamental length scale of the Abelian Laughlin state at the same filling. 

In Fig.~\ref{fig2}(f) the density profiles indicate that the size of the quasi-hole of the Moore-Read state is comparable or larger than that of the Laughlin-1/2 state. The fact that the Abelian Moore-Read quasihole subspace actually has a smaller fundamental length scale could be a manifestation of the non-Abelian nature of the Moore-Read state: unlike the $\nu=1/2$ Laughlin state after a single flux insertion, the Abelian Moore-Read quasihole of charge $e/2$ is only a \emph{subspace} of the zero energy manifold (i.e. the Moore Read CHS), since the single flux can also fractionalize. This could imply the Moore-Read quasiholes are more sensitive to impurities in the systems as we will show next.

\textit{Effective disorder potentials in realistic systems--} The vNL Hamiltonian not only captures the fundamental length scale of the CHS and their corresponding fractional quantum Hall phases, but it also helps understand how quasiparticles of such topological phases are affected by the  disorder potentials in the system.  Although screened Coulomb interactions from impurities induce strong long-wavelength disorder potentials, short-wavelength disorder can also be significant, as observed in GaAs experiments\cite{wavelength1,wavelength2,wavelength3}. Our analysis from previous sections shows that the effect of the local potential lattice is quickly suppressed when its spacing decreases below such length scales. Thus, for disorder potentials in the 2D system from impurities, the short wavelength part should also have negligible effects on the dynamics of quasiholes. 

More explicitly we compute the effect of periodic one-body potential $V(r)=cos(2\pi t_x x)cos(2\pi t_y y)$, where $t_{x(y)}$ is the wavelength along x or y directions, playing the same role as spacing in the case of the local potential lattice. Similar to the delta potential lattice, one can see the bandwidth of a periodic potential decays rapidly as the wavelengths decrease as shown in Fig. 3(a). When the wavelength of the one-body potential is close to or smaller than the magnetic length, its effect on the dynamics within the LLL is negligible given the very small bandwidth. The same can be generalized to fractional quantum Hall phases. As long as the one-body potential strength is much smaller than the incompressibility gap, and the wavelength is smaller than the fundamental length scale given by the vNL Hamiltonian, its effect on the dynamics of the anyon liquid made of a collection of the quasiholes is also negligible.

\begin{figure}[!h]
\centering
\includegraphics[width=1.0 \linewidth]{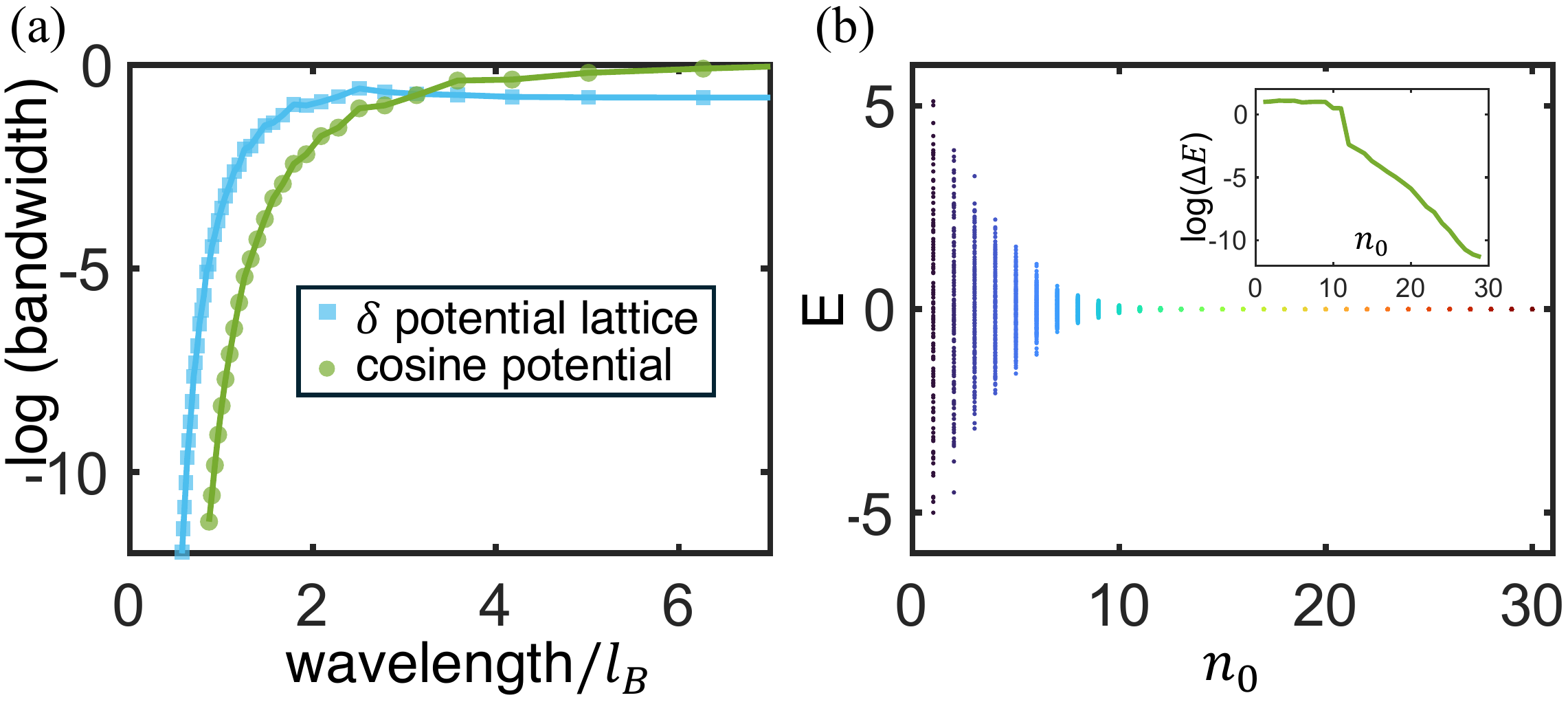}
\caption{(a). Bandwidths' logarithms of delta potential lattice (square cyan points) and cosine potential (circular green points). (b). The spectra of short-wavelength terms of a random potential. The inset figure is the logarithm of energy difference summation between the random potential and its long-wavelength terms after truncation from $n_0$.}
\label{fig3}
\end{figure}

If we start with a random disorder potential mimicking the situation in realistic samples, it can be decomposed into Fourier series: $V(x,y)=\sum_{n_x,n_y=-\infty}^{\infty} c_{n_x, n_y}e^{i 2\pi (\frac{n_x x}{l_x}+\frac{n_y y}{l_y})}$, as a sum of periodic potentials with wavelength $l_{x(y)}/n_{x(y)}$, where $n_x$ and $n_y$ are integers,$l_{x(y)}$ gives the size of the torus.  We calculate the energy spectrum of the random potential $V$, then truncate its short-wavelength terms by keeping only terms with $|n_{x(y)}|\leq n_{0}$. The spectrum of the truncated part of the potential has negligible bandwidth when $n_0$ is large enough (corresponding to wavelength smaller than the fundamental length), see Fig.~\ref{fig3}(b), so they do not affect the dynamics of the (quasi)hole excitations, as quantified by the summation of energy difference in the inset. As a generalization, the dynamics of anyons in FQH states will be also robust against the disorder potentials, more so than the IQH counterpart because the fundamental lengths of the FQH phases are larger. High-order perturbations arise due to finite incompressible gaps in realistic systems. However, when disorder wavelengths (or superlattice spacings) are smaller than the fundamental length, such Hamiltonian is very close to the identity matrix, rendering the effect of LL mixing on bandwidth and fundamental length negligible, as shown in detail in the supplementary materials\cite{supple}.

\textit{Summary and discussion--}In summary, we show that the continuous translational invariance can be surprisingly robust within a single LL, or the CHS of the FQH phases, in the presence of the superlattice potential. With this, we can define the vNL Hamiltonian as a lattice of periodic one-body delta potentials that gives the fundamental length scale of different quantum Hall phases, that physically characterizes the length scales of the density modulation of the anyonic excitations in real space and their dynamics. For small disorder that does not close the incompressibility gap (i.e. no phase transition), it can still affect the dynamics of the anyons within the same topological phase, including the pinning of the anyons and the deformation of the anyonic shapes\cite{statistics3}. We show here that the large wave vector parts of disorder have negligible effects on the dynamics of anyons in realistic systems, and anyons from FQH phases at low filling factors are thus less sensitive to the disorder potentials in the system.

The study of fundamental length and the superlattice effects inspired the finding of rather surprising suppression of Anderson localization in Landau levels\cite{nilanjan}. Furthermore, the fundamental length is relevant to enabling anyonic braiding and exhibiting their exotic quantum statistics. To realize the anyons braiding while avoiding compromising the accuracy of exchange statistics and non-topological effects, the spatial separations among braiding paths must be large and the fundamental length gives the relevant length scale. The potential relationship of fundamental length and the band quantum metric is interesting given that they behave differently in different Landau levels, warranting further in-depth studies.

\textit{Acknowledgement--} We would like to acknowledge useful discussions with Ha Quang Trung and Yuzhu Wang. This work is supported by the NTU grant for the National Research Foundation, Singapore under the NRF fellowship award (NRF-NRFF12-2020-005), and Singapore Ministry of Education (MOE) Academic Research Fund Tier 3 Grant (No. MOE-MOET32023-0003) Quantum Geometric Advantage.

\onecolumngrid
\pagebreak
\widetext
\begin{center}
\textbf{\large Supplementary materials for ``Robust translational invariance in Landau Levels against lattice potentials and disorders''}

\end{center}
\setcounter{equation}{0}
\setcounter{figure}{0}
\setcounter{table}{0}
\setcounter{page}{1}
\makeatletter
\maketitle
\renewcommand{\theequation}{S\arabic{equation}}
\renewcommand{\thefigure}{S\arabic{figure}}
\renewcommand{\thetable}{S\arabic{table}}
\renewcommand{\bibnumfmt}[1]{[S#1]}
\renewcommand{\citenumfont}[1]{S#1}

\section{A. Bandwidth of square superlattice on the disk and torus}

In the supplementary material, we give the derivation of square lattice bandwidth on the disk in the LLL, and show the acutely decreasing bandwidth in the LLL on the torus.

The Hamiltonian given by $\delta$ functions can be written as:
 \begin{equation}
     H=\sum_{a,b} \delta (x-at)\delta (y-bt),
 \end{equation}
 where $(at,bt)$ are the positions of the delta potentials, and $t$ is the spacing of the square lattice, $a$ and $b$ are integers. The projection to LLL of such local potential lattice can be given by spanning of a series of coherent states $\ket{X}=e^{i\hat{R^a}X_a}\ket{0}$:
 \begin{equation}
     H'=\frac{1}{2\pi}\sum_{X}\ket{X}\bra{X}= P_{LLL} \sum_{a,b} \delta (x-at)\delta(y-bt) P_{LLL},
 \end{equation}
where $P_{LLL}$ is the projection operator to the LLL: $P_{LLL}=\sum_{m=0}^{m=\infty}\ket{m}\bra{m}$, where $\ket{m}$ constitute the basis of LLL and m is the index of angular momentum. In the LLL, the real space expression of such Hamiltonian is:
\begin{equation}
\begin{split}
    \bra{x,y}H'\ket{x,y}
    &=\frac{1}{2\pi}\sum_{a=-\infty,b=-\infty}^{\infty} \frac{1}{2\pi}e^{-\frac{1}{2}\left ((at+x)^2+(bt+y)^2\right )}\\
    &=\frac{1}{2\pi} \cdot \frac{\theta\left[3,\frac{x\pi}{t},e^{-\frac{2\pi^2}{t^2}}\right]\cdot \theta\left[3,\frac{y\pi}{t},e^{-\frac{2\pi^2}{t^2}}\right]}{t^2},\\
\end{split}
\end{equation}
where $\theta[3,u,q]$ is the third elliptic theta function, defined as:
\begin{equation}
    \theta[3,u,q]=1+\sum_{n=1}^{\infty} q^{n^{2}}cos(2nu)
\end{equation}
For the theta function $\theta[3,u,q]$, it also has the other form:
\begin{equation}
    \theta[3,u,q]=\prod_{m=1}^{\infty}\left(1-q^{2m}\right)\left(1+q^{2m-1}\omega^2\right)\left(1+q^{2m-1}\frac{1}{\omega^2}\right),
\end{equation}
where $\omega=e^{\imath u}$.
In order to find the maximum and minimum point of some spacing, the determinant term should be $\left(1+q^{2m-1}\omega^2\right)\left(1+q^{2m-1}\frac{1}{\omega^2}\right)$. Then 
\begin{equation}
\begin{split}
    \left(1+q^{2m-1}\omega^2\right)\left(1+q^{2m-1}\frac{1}{\omega^2}\right)&=1+q^{2m-1}\left(1+\omega^2+\frac{1}{\omega^2}\right)\\
    &=1+q^{2m-1}\left(1+e^{2\imath u}+e^{-2\imath u}\right)\\
    &=1+q^{2m-1}\left(1+2cos(2u)\right)\\
\end{split}
\end{equation}
One can see that each term of the product is positive and when $u=0$ it takes the maximum while when $u=\frac{\pi}{2}$ it takes the minimum. It means that when $x=0+ct$ and $y=0+dt$ it reaches maximum, and when $x=\frac{t}{2}+ct$ and $y=\frac{t}{2}+dt$ it reaches the minimum with $c$ and $d$ as integers. Note that $\Delta E$ as the difference of the maximum and minimum, it is also the bandwidth of Hamiltonian $H$ in Eq. (S2).

Defining that $q=e^{-\frac{2\pi^2}{t^2}}$,
\begin{equation}
    \begin{split}
        \Delta E& = \frac{1}{2\pi t^2}\left(\theta\left[3,0,e^{-\frac{2\pi^2}{t^2}}\right]\cdot \theta\left[3,0,e^{-\frac{2\pi^2}{t^2}}\right] - \theta\left[3,\frac{\pi}{2},e^{-\frac{2\pi^2}{t^2}}\right]\cdot \theta\left[3,\frac{\pi}{2},e^{-\frac{2\pi^2}{t^2}}\right]\right)\\
        & = \frac{1}{2\pi t^2}\left[\left(\sum_{n}q^{n^2}\right)\left(\sum_{m}q^{m^2}\right) - \left(\sum_{n'}(-1)^{n'}q^{n'^2}\right)\left(\sum_{m'}(-1)^{m'}q^{m'^2}\right)\right]\\
        & = \frac{1}{2\pi t^2}\left(\sum_{n,m}q^{m^2+n^2}-\sum_{n',m'}(-1)^{n'+m'}q^{m'^2+n'^2}\right)\\
        & = \frac{2}{2\pi t^2}\left(\sum_{n\in odd}q^{n^2}\sum_{m\in even}q^{m^2}+\sum_{n'\in even}q^{n'^2}\sum_{m'\in odd}q^{m'^2} \right)\\
        & = \frac{4}{2\pi t^2} \theta[2,0,q^4]\cdot \theta[3,0,q^4]\\
        & = \frac{2}{\pi t^2} \theta\left[2,0,e^{-\frac{8\pi^2}{t^2}}\right]\cdot \theta\left[3,0,e^{-\frac{8\pi^2}{t^2}}\right],\\
    \end{split}
\end{equation}
where $\theta[2,u,q]$ is the second ellptic theta function, given by 
\begin{equation}
    \theta[2,u,q]=2q^{\frac{1}{4}}\sum_{n=0}^{\infty}q^{n(n+1)}cos((2n+1)u)
\end{equation}

\begin{figure}[!h]
\centering
\includegraphics[width=0.35 \linewidth]{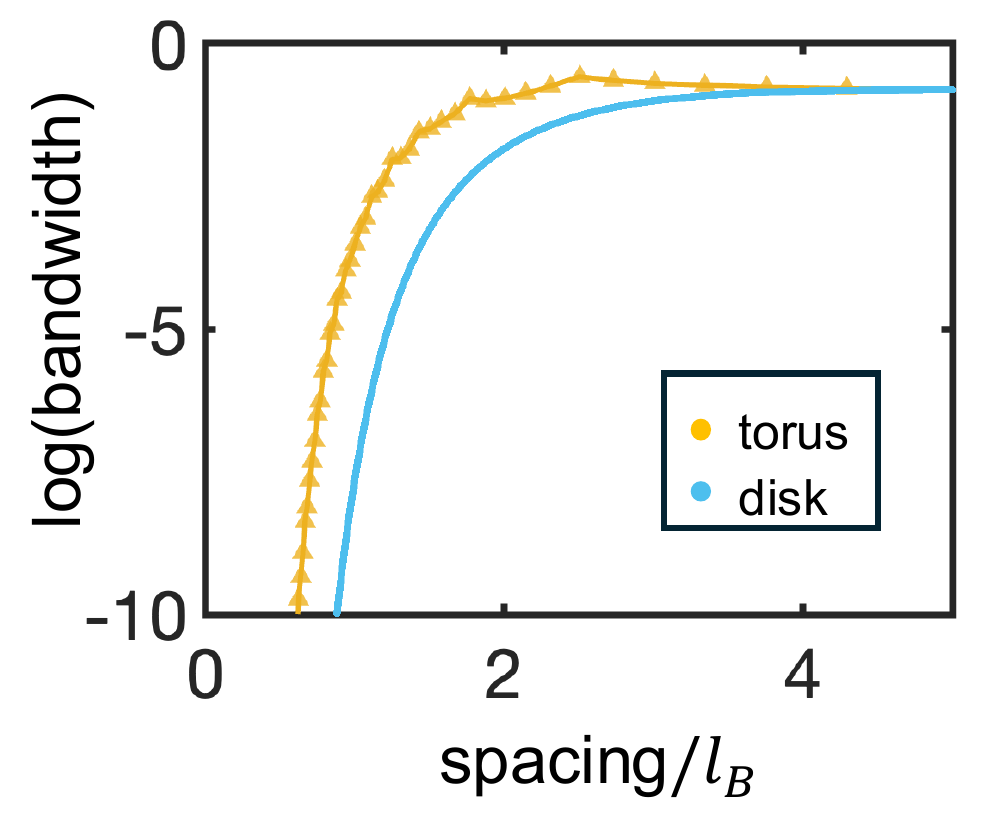}
\caption{Logarithm of bandwidth of square lattice potentials on the torus(yellow dots and line) and disk(cyan line) in the LLL.}
\label{fig6}
\end{figure}

Such bandwidth sharply reduces as the superlattice spacing decreases, this is valid not only on the infinite plane but also for the periodic geometry, torus, as shown in Fig. S1. Even though the cutoff on the torus is much higher than disk, the narrow bandwidth implies the translational invariance for superlattice with small spacing. 

\section{B. Fundamental lengths dependence on bandwidth cutoff and constant ratio}

As long as we define the fundamental length within the LLL to be the magnetic length (which is how we choose the bandwidth cut-off depending on the detailed form of the periodic potential), the ratio of the fundamental length in different topological phases is not affected by bandwidth cutoff benchmark or the particular form of the periodic one-body potential. Below we compare the behavior of the square lattice potential with that of a periodic step function as an example. The periodic step function is given by below, as shown in Fig. S2(a).

\begin{equation}
    V(x,y)=f(x)g(y),
\end{equation}
in which 

\begin{equation}
    f(x)=\left\{\begin{array}{rcl}
        \frac{1}{2}, &  a \cdot t \leq x < \left( a+\frac{1}{2}\right) \cdot t\\
        -\frac{1}{2}, & \left( a+\frac{1}{2}\right) \cdot t \leq x < (a+1) \cdot t
    \end{array}
    \right . ,
\end{equation}
\begin{equation}
    g(y)=\left\{\begin{array}{cc}
        \frac{2}{3}, &  a \cdot t \leq y < \left( a+\frac{1}{2}\right) \cdot t\\
        -\frac{1}{3}, & \left( a+\frac{1}{2}\right) \cdot t \leq y < (a+1) \cdot t
    \end{array}
    \right. ,
\end{equation}
in which $a$ is an integer and $t$ is the period of the potential.

\begin{figure}[!h]
\centering
\includegraphics[width=0.9 \linewidth]{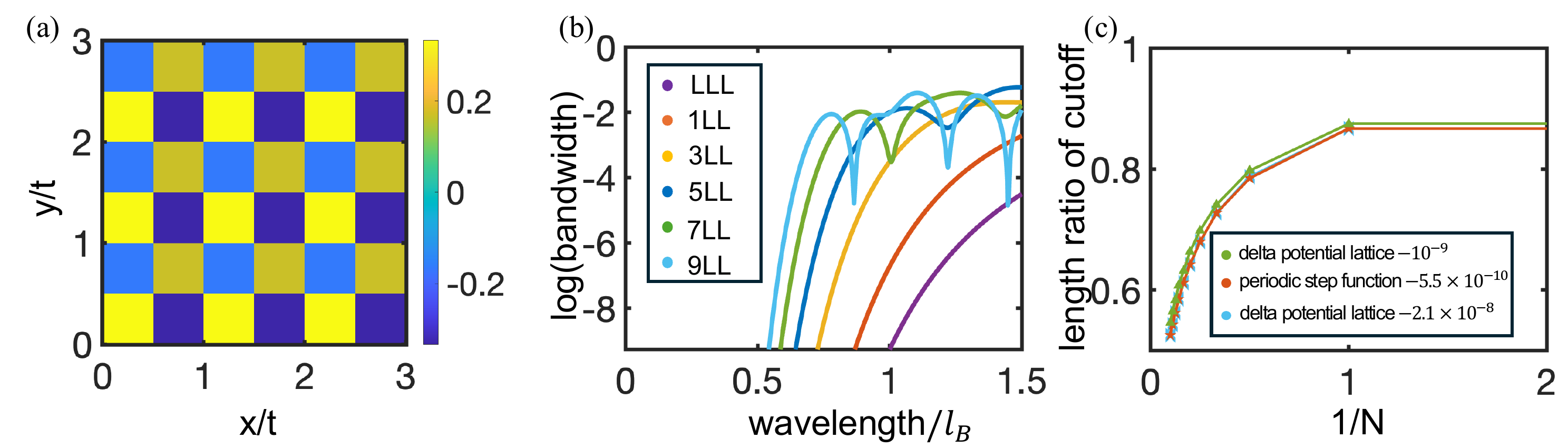}
\caption{ (a). The periodic step function in real space. (b). Bandwidth logarithm of the periodic step function on the disk in different LLs. (c). The fundamental lengths given by local potential lattice in the different LLs when the cutoff is $2.1\times 10^{-8}$ as well as $10^{-9}$, in which $N$ is the index of LL, and that of the step function corresponding cutoff $5.5\times 10^{-10}$.}
\label{fig4}
\end{figure}

Fig. S2(b) shows the bandwidth logarithm of the periodic step function as a function of wavelength on the disk and that of square lattice potential is shown in Fig. 2(a) in main text as comparison. Smaller than the square lattice potential bandwidth, the bandwidth of the periodic step function in the LLL is about $5.5\times10^{-10}$ when wavelength is $\ell_B$. Considering such energy scale as benchmark, the fundamental lengths for higher LLs are given, as shown in Fig. S2(b). The slight difference between the two bandwidth logarithm figures and the distinction of bandwidth cutoffs imply the bandwidth dependence on particular form of the periodic one-body potential. However, Fig. S2(c) shows the highly agreement of the fundamental length ratios of high LLs to the LLL for both the periodic step function (red dots and line) and the local potential lattice(cyan dots and line), implying the robustness of fundamental length ratios against the particular form of periodic one-body potential. Therefore, we conclude that while the choice of periodic potential influences the spectral structure and the bandwidth scale—for instance, the cosine potential can produce cutoffs as small as $10^{-17}$, it does not affect the scale of the fundamental length.

Furthermore, we also emphasize only the ratio of the fundamental lengths in different phases is physically important, which should be insensitive to the choice of the cut-off as long as it is small enough. Taking the square lattice potential as an example, if we take the bandwidth benchmark as $10^{-9}$, the corresponding fundamental length in the LLL is about $0.93\ell_B$. And the fundamental length ratios between high LLs and the LLL under such benchmark are shown by the green triangular points and line in Fig. S2(c). One can see that the ratio of the fundamental lengths in high LLs to that of LLL remain more or less the same as the case of bandwidth benchmark $2.1\times 10^{-8}$ , as shown by the great coincidence of the green and blue lines and dots.

In the main text, we choose $2.1\times10^{-8}$ as the cutoff benchmark for square superlattice due to some physical consideration: in the LLL, this cutoff gives the fundamental length as the magnetic length, the only microscopic length scale in the system. In this way, we define the fundamental length of the LLL as equal to $\ell_B$, thus fixing the overall constant factor. 

\section{C. Negligible high-order perturbation effect}

In order to construct the conformal Hilbert space, the scale of the incompressible gap is the main energy scale, and we consider it as infinity comparing with the superlattice disorder potential.  And the effect of disorders or one-body potentials play a role of first order perturbation.  Theoretically this allows us to focus on the fundamental properties of the gapped topological phases (e.g. the emergence of robust translational invariance) and the extraction of this new length scale characterizing different topological order. After all, even the Hall conductivity is only strictly quantized/defined in the limit of infinite gap, and that is the similarly the case in this work. However, in realistic systems the incompressible gap is always finite thus the high-order perturbations emerge. But for the the superlattice potential or the disorder potential with wavelength(lattice spacing) smaller than the fundamental length, the effect of high-order perturbations to the bandwidths and fundamental lengths is insignificant, as we show below taking LL mixing as an example.

The calculation is done on the torus with $N_{\Phi}$ fluxes in a single LL (analogous results can be obtained on disk geometry), where we consider the Hamiltonian:
\begin{equation}
    H=H_{vNL}+H_{k},
\end{equation}
in which $H_k$ is the kinetic Hamiltonian. Due to the large kinetic energy, the total energy spectra are separated into several groups, each of them corresponds to a single LL. And we focus on the bandwidth of lowest $N_{\Phi}$ energies, which are the spectra of the dressed LLL after corrections from LL mixing. Our calculation explicitly includes perturbations of all orders, not just the second order perturbation. We thus compare them with the spectra obtained from calculations solely within the LLL. The discrepancy between the two serves as a quantitative measure of high-order perturbation effects. 

To illustrate this issue more explicitly, we tune the cyclotron gap strength and give how the bandwidth and the fundamental length of the LLL are affected by the LL mixing. Fig. S3(a) shows the bandwidth of the dressed LLL, when the full Hilbert space includes both the LLL and the 1LL. $\alpha$ denotes the ratio between the LL gap and the strength of the local potential, and the case $\alpha=\infty$ corresponds to the ideal case in which the incompressible gap is infinitely large and only the LLL is considered (the undressed LLL).
Fig. S3(b) shows bandwidth of the dressed LLL, when the full Hilbert space consists of the lowest $N$ LLs at $\alpha=1$, implying that the scale of LL gap is on the same order as the local potential strength. One can see  the high coincidence of superlattice bandwidths between the dressed LLL and the LLL itself indicates the insignificant impact of LL mixing. Table. S1 and Table. S2 list the percentage change of fundamental length, with the fundamental length on the single LLL as $\ell_B$. Even when the incompressibility gap (here the cyclotron energy) is as small as the superlattice potential ($\alpha=1$), the change of the bandwidth due to LL mixing and thus the change in the extracted fundamental length is extremely small. For the realistic case where the cyclotron gap should be much larger than the superlattice potential or the impurity/disorder potential, their effect is even weaker.

\begin{figure}[!h]
\centering
\includegraphics[width=0.6 \linewidth]{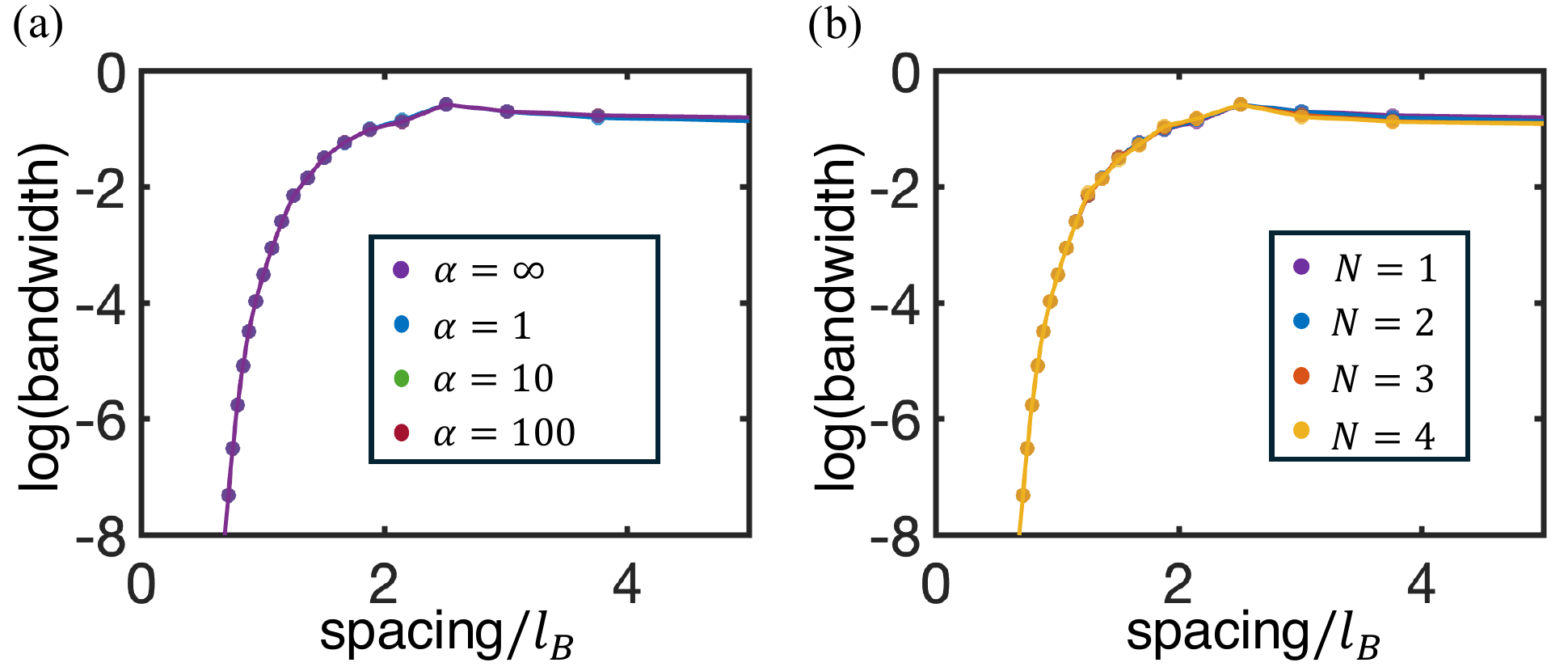}
\caption{Logarithm of bandwidths of square delta potential lattice on the torus of the dressed LLL. (a). The bandwidths logarithm when the full Hilbert space includes both the LLL and the 1LL corresponding to different $\alpha$, which is the ratio of kinetic energy and local potential strength. (b). The bandwidths logarithm of dressed LLL, when full Hilbert space consist of N lowest LLs, taking $
\alpha$ as 1.}
\label{fig4}
\end{figure}

\begin{table}[!htbp]
    \centering
    \renewcommand{\arraystretch}{1.5}
    \begin{tabular}{| c | c | c | c | c |}
    \hline
         $\alpha$ & $\infty$ & 1 & 10 & 100 \\
         \hline
         fundamental length percentage change(\%) & 0 & $3\times 10^{-6}$ & $3\times 10^{-7}$ & $1\times 10^{-8}$  \\
         \hline
    \end{tabular}
    \caption{Percentage change of fundamental length on the dressed LLL due to the mixing with the 1LL, for different cyclotron gap strength.}
    \label{Table1}
\end{table}

\begin{table}[!htbp]
    \centering
    \renewcommand{\arraystretch}{1.5}
    \begin{tabular}{| c | c | c | c | c |}
    \hline
         mixed LL number $N$ & 1 & 2 & 3 & 4 \\
         \hline
         fundamental length percentage change(\%) & 0 & $3\times 10^{-6}$ & $1.5\times 10^{-4}$ & $1.5\times 10^{-4}$  \\
         \hline
    \end{tabular}
    \caption{Percentage change of fundamental length on the dresses LLL due to the mixing with the N lowest LLs, with $\alpha=1$.}
    \label{Table2}
\end{table}

As soon as the superlattice constant is small compared to the magnetic length, the bandwidth and change of fundamental length are always extremely small. This is in some sense expected since the fundamental length characterizes topological order and is thus rather robust against perturbation. And the extremely small bandwidth implies that the superlattice potential Hamiltonian is very close to the identity matrix (thus not breaking the continuous magnetic translational invariance). This implies the matrix elements between single particle states between different LLs are also very small (up to on the order of the bandwidth or $10^{-8}$).

Thus, even though with a finite incompressible gap, the high-order perturbation effect of superlattice or disorder (with wavelength smaller than the fundamental length) on the dynamics of anyons (or low energy excitations below the incompressibility gap) is highly suppressed, due to the robust translational invariance in such truncated Hilbert space against the one-body potentials.

\end{document}